# SigniFYI-CDN: merged communicability and usability methods to evaluate notation-intensive interaction

Juliana Soares Jansen Ferreira[1,2], Clarisse Sieckenius de Souza[1],
Rafael Rossi de Mello Brandão[1,2] and Carla Faria Leitão[1]

Semiotic Engineering Research Group (SERG) [1]
Departamento de Informática, PUC-Rio
{clarisse, cfaria}@inf.puc-rio.br

IBM Research Brazil [2]
{jjansen, rmello}@br.ibm.com

## ABSTRACT

We present **SigniFYI-CDN**, an inspection method built from previously proposed methods combining Semiotic Engineering and the Cognitive Dimensions of Notations. Compared to its predecessors, **SigniFYI-CDN** simplifies procedural steps and supports them with more analytic scaffolds. It is especially fit for the study of interaction with technologies where notations are created and used by various people, or by a single person in various, and potentially distant, occasions. In such cases, notations may serve several purposes, like (mutual) comprehension, recall, coordination, negotiation, and documentation. We illustrate **SigniFYI-CDN** with highlights from the evaluation of a computer tool that supports qualitative data analysis. Our contribution is a simpler tool for researchers and practitioners to probe the power of combined communicability and usability analysis of interaction with increasingly complex data-intensive applications.

## INTRODUCTION

Interaction where the creation and manipulation of notations is critical for the user's activities is not new, but has gained importance in notation-intensive technologically-enabled domains such as social tagging, end user development, and data analysis, for example. Most of the research on the usability of notations in human-computer interaction (HCI) has been carried out using the Cognitive Dimensions of Notations (CDN) framework [1], a long standing territory of research about cognitive costs associated with the human use of computer representations [2]. In 2012, as part of our research with Semiotic Engineering [3], we began to explore the combination of usability inspections using CDN with communicability inspections using semiotic methods [4,5,6]. In 2016, our Semiotic Engineering book presenting a suite of tools for the study of human-centered software development [7] incorporated CDN to half of the proposed suite's modules.

Semiotic Engineering and CDN have important similarities and distinctions. Starting with similarities, in both approaches user interface representations are the key object of interest for HCI design and evaluation. Semiotic Engineering refers to them as 'signs', whereas CDN refers to them as 'notations'. Nevertheless, the two approaches have fundamental distinctions. CDN, following a classical user-centered perspective, focuses on *users* and the activities that they may or must carry out using interface notations. Semiotic Engineering, following the seminal perspective of pioneers such as Andersen [8] and Nadin [9], views HCI as a particular case of computer-mediated human communication, between those who create and those who use computer technologies. Their communication is mediated by systems interfaces, which are *proxies* of their creators, that is, they speak for their creators at interaction time, by means of typically non-verbal interaction signs such as mouse clicks, visual object manipulations, gestures, as well as command lines and programming scripts. The combination of CDN and Semiotic Engineering thus allows us to expand the scope of each individual approach, and to articulate how social communication and individual cognition aspects of interaction can affect and change the quality of the users' interaction with computer technologies.

The combination of communicability and usability analysis can help us face increasingly complex interaction design challenges regarding the meaning of computer signs and notations [7]. However, the power of such combination risks to be missed in the original version of tools in the SigniFYI Suite (pronounced as 'signify suite') for two main reasons.

Firstly, the combined methods have been published in a book for readers interested in semiotic contributions to software design and engineering, framed as a human-centered computing activity. Secondly, and more importantly for the research presented in this paper, the SigniFYI Suite can be daunting at first encounter. Researchers interested in understanding and exploring the proposed combination of techniques – to use them, to critique them, or to improve them – may be discouraged by the initial investment required to use them effectively. As a consequence, an entire branch of investigation may fail to develop, a risk that we propose to attenuate with our research contribution.

This paper presents a new semio-cognitive inspection method called **SigniFYI-CDN**. It is the result of incremental research that takes portions of the SigniFYI Suite, selects a small set of core Semiotic Engineering and CDN concepts, defines straight-forward inspection procedures, and provides practical scaffolds to support beginners. Our aim is to contribute to deepen and sharpen the debate about non-cognitive approaches to HCI evaluation, by allowing debaters to try effectively this particular combination of perspectives and see what it can do.

Although **SigniFYI-CDN** can be used to evaluate interaction design in general, it is especially fit for the study of interaction with computer technology where notations are created and used by various people, or created by a single person but used in various and potentially distant occasions. In these cases, notations may serve one or more of several purposes, like supporting (mutual) comprehension, recall, coordination, negotiation, documentation, and decision making. We believe that **SigniFYI-CDN** is one of several emerging instances of a new breed of methods to evaluate notation-intensive interaction, such as that which originates massive amounts of data used by contemporary intelligent systems or by expert quantitative and qualitative data analysts.

We illustrate the power of our method with highlights from the evaluation of interaction with QDA Miner Lite©, a qualitative data analysis tool [10]. The inspection scenario is focused on coding – a notation-intensive task par excellence – in the context of a systematic literature review.

The paper is structured in five sections, starting with this brief introduction. In the next section, we comment on related work. The third and fourth sections present the gist of the paper: a description of the method; and illustrative excerpts of QDA Miner Lite© evaluation. Finally, the last section concludes the paper and points at future work.

## RELATED WORK

Evaluation methods comprise a vast portion of research in HCI to-date. We focus our commentary on related work that, by contrast with ours, points at why **SigniFYI-CDN** is a relevant contribution at this stage of evolution in HCI. We cover essentially work that has been explicitly proposed to analyze computer notations designed for human use, and work that explicitly complements or contrasts usability evaluation with (some level of) semiotic analysis. Finally, to clarify the importance of having appropriate methods for notation-intensive kinds interaction, we briefly comment on previous work about the evaluation of programming languages and information visualizations, which are both notation systems *par excellence*.

The most widely known and used approach to the evaluation of notations in HCI is CDN [1,11]. According to its proponents, CDN "describes necessary (though not sufficient) conditions for usability, deriving usability predictions from the structural properties of a notation, the properties and resources of an environment, and the type of activity." [1] CDN-based evaluation requires interpretive analysis, that is, an evaluator's interpretation, systematization and decision of what the presence or absence of cognitive characteristics in notations mean in terms of cognitive loads imposed to their users. Examples of CDN evaluation in notation intensive interaction include computer programming [12,13], information visualization [14], and collaboration support [15].

An alternative framing of notational analysis is Moody's Physics of Notations (PoN) [16]. PoN is specifically proposed as a theoretical basis for the construction of visual notations used in software engineering. Compared to CDN, the scope of PoN is narrower. However, despite its name, Moody's approach essentially integrates linguistic and semiotic dimensions with cognitive ones. The result is a complex theoretical framework, which has required operationalization for practical use [17].

Moody's integration of semiotic dimensions in the analysis of computer languages was first proposed by Zemanek in 1966 [18], followed by Nadin [9] and Andersen [8]. In HCI, semiotic methods to evaluate the communicability of interaction design appeared in 2000 and the following decade [19,20,21], proposed by researchers working with Semiotic Engineering. Another breed of semiotic methodology was proposed and evolved by researchers working with Organizational Semiotics [22,23]. None of the early Semiotic Engineering or Semiotic Organization methodologies provide explicit integration with cognitive dimensions of HCI evaluation, although they all mention significant relations between semiotic and cognitive aspects of HCI. Such is also the case of other semiotic studies that do not specifically propose (or have yet evolved into) HCI evaluation methods (e.g. [24,25,26]). Recent work in formal methods for Semiotic Engineering [27] incorporates sensory dimensions to the analysis, but uses formal verification and model checking techniques to assess the efficiency and effectiveness of metacommunication.

The first known approach to integrate communicative and cognitive dimensions with one another is Hundhausen's [28]. The author proposed to extend CDN with four semiotic dimensions (which he called 'communicative dimensions'), motivated by the rise of visual programming languages. This work was followed by further research on the mediating role of notations in collaborative work [15]. The larger conceptual and methodological integration in semio-cognitive analysis, however, appeared in association with our own Semiotic Engineering research [4,5,6,29], with research frameworks and analysis that explicitly articulated Semiotic Engineering with CDN. This approach has been recently consolidated into a suite of concepts, methods and capture & access infrastructure model – the SigniFYI Suite [7].

In spite of its promise and depth of theoretical articulation, the SigniFYI Suite presents considerable challenges for researchers who are not initiated in Semiotic Engineering. An example of such challenges is the need to master the Semiotic Engineering classification of metacommunication signs and communicability breakdowns. Yet, the opportunity for semio-cognitive integration has been underlined by researchers working in other hot areas of research. One of them Information Visualization [30,31], where visual signs play critical role in supporting communication and comprehension. The other area is the study of application programming interfaces (API). The role of APIs in contemporary software development practices calls for more usable APIs. Yet, API designers need further support to communicate their expectations and intentions regarding key programmatic features of their design [32]. One of the recently emerging ideas is to turn APIs into more conversational artifacts, using a combination of communicability and usability perspectives [33,34].

In view of such previous work, **SigniFYI-CDN** does with the SigniFYI Suite [7] a similar operationalizing job as Störrle and Fish [17] do with Moody's PON [16]. And by so doing, it creates new possibilities for Information Visualization and API Design researchers, too.

## THE METHOD

**SigniFYI-CDN** integrates *communicability* and *usability* evaluation into a single process. **Communicability** is the artifact's ability to communicate, on behalf of those who created it, its design principles, purposes, value, and modes of operation, in pragmatically adequate ways. Communication is achieved as interaction unfolds. Pragmatic adequacy is the result of adopting communication strategies that maximize the chances of achieving all the effects intended by communicators. **Usability**, in this method's particular context, is centered on cognitive aspects of notations. Therefore, non-cognitive aspects of usability, such as the user's satisfaction and productivity, are not directly addressed.

**SigniFYI-CDN** consists of four procedural blocks: Preparation; Semiotic Analysis; Cognitive Analysis; and Articulation. They are logically ordered, but the second and third can be executed either in sequential mode (*i.e.* Semiotic Analysis is *completed* before Cognitive Analysis begins), or interleaved mode (*i.e.* Semiotic Analysis is *started* before Cognitive Analysis begins, both blocks alternating portions of analysis until the process is completed for both). The precedence of the second block over the third one means that the designer-user communication mediated by the artifact's interface governs the nature and purposes of the user's cognitive tasks.

**Preparation** is a fairly standard procedure in HCI inspection methods (e. g. [35]. Inspectors must be thoroughly familiar with the profile of users for whom they advocate with their analysis, study the artifact that they are about to analyze, and define the inspection baseline. This baseline comprises the user profile, the inspection scenario, and the mental and physical tasks required to run through the scenario. Notice that only the interactions involved in running the scenario – including variations around them, with which the inspector explores multiple alternative paths that users might follow – are analyzed.

**Semiotic Analysis** begins with a systematic examination of two kinds of communication source, keeping the focus on the inspection baseline (user profile, scenario, and tasks):

- Content of the inspected artifact's official website, online help and documentation, user manuals and guides, tutorials, or other available source of information, explanation, demonstration and illustration;
- Communication content, patterns and strategies used in static interface representations, as well as in dynamic interaction flows (including alternatives), with special attention to: ***who*** (system, user or users) says ***what*** (types of possible message content), ***to whom*** (system, user or users), ***how*** (types of representations), ***when*** (types of channels and controls), ***where*** (contexts), ***why*** (assumptions that explain/justify communication), and ***what for*** (effects that communication should achieve).

During the examination, inspectors register noteworthy communicability features of the artifact. 'Noteworthy' features depend on the purposes of inspection. For example, if the inspector's purpose is to compare design alternatives, good and bad communicability features that can be weighed against each other will be 'noteworthy'. Likewise, if the purpose is to improve an existing design, 'noteworthy' features can concentrate on problems and issues, with smaller emphasis on successful communication.

The Semiotic Analysis ends with a progressive tabulation of findings. In Figure 1, we show the structure of tabulation for the lower level of abstraction in the progression. The inspector checks the results of his Semiotic Analysis of mediated designer-user communication and decides if they are compatible with the inspection baseline, in view of related noteworthy communicability features.

| Inspection Baseline: | Compatible with Inspected Designer-User Communication? | | Noteworthy Communicability Features |
|---|---|---|---|
| | Yes | No | |
| the User's profile | | | |
| the User's goals | | | |
| the User's needs | | | |
| the User's preferences | | | |
| the System's description | | | |
| the System's functionality | | | |
| the System's mode of use | | | |
| alternative modes/purposes of system's use | | | |

**Figure 1. Structuring Chart for Conclusions at the Lower Level of Abstraction in Semiotic Analysis**

As an example of what the chartered conclusions at this level may be, suppose that the baseline states that the profiled user is an undergraduate student of an introductory course on qualitative methods of analysis, who has just had his first class on *coding* (the systematic categorization of data based on principled and justifiable interpretation of what the data *means*). His assignment for next class is to use a freely distributed computer-aided qualitative data analysis (CAQDAS) tool called QDA Miner Lite to: (1) code a presidential speech that his teacher has selected for the exercise; (2) write a 3-page report with his analysis, including the codes he used (followed by examples of coded passages), quantitative graphics showing the frequency of codes in the material, and finally his interpretation of what his coding *means* with respect to the president's positioning in view of events reported in the analyzed speech.

In Figure 2, we show examples of static interface and dynamic interaction signs that the inspector finds while running through the possible scenario paths and executing associated tasks. The student (profiled user) can reach the 'Coding Frequency' dialog depicted in '1' (upper half of the image), where one of the icons in the toolbar explicitly communicates that the [Coding Frequency] table can be added to the [qualitative analysis] report. This is exactly what the student needs and wants to do.  Yet, when he clicks on the icon, the message he gets is shown in '2' (lower half of the image): "This feature is only available in full version!".

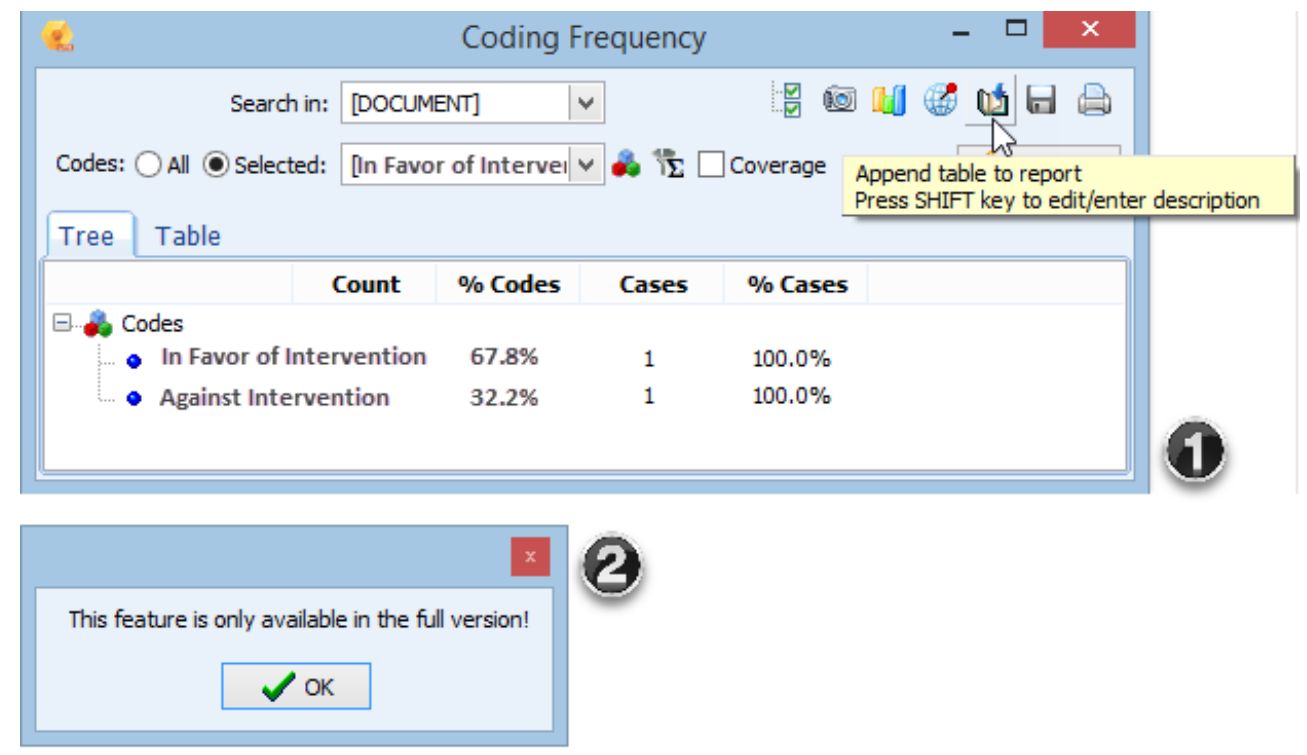

**Figure 2. QDA Miner Lite Interface and Interaction Signs**

The inspector can add this noteworthy communicability feature on the table depicted in Figure 1. It is clearly related to 'the user's goals'. Interface signs firstly suggest that the user *can* include coding frequency information in a report, but as soon as the user communicates that this is what he needs the system replies that this cannot be done with the current version of the software (only with the full version). This apparently negative communicability feature is also associated with the 'user's needs' (data analysis is generally carried out to be presented to others, by means of a technical document or report), which are not fully contemplated with QDA Miner Lite. This feature can additionally be associated with 'system's description' and 'system's functionality' factors, inasmuch as interactive messages about what the system *does* are not consistent.

At this lower level of abstraction, the conclusion of Semiotic Analysis is hardly more than a semiotic counterpart of typical usability analysis conclusions that can be achieved with CDN or other methods (*e. g.* a Cognitive Walkthrough [35]). At the next, middle level of abstraction, the inspector must consider three factors in mediated communication: the expression of the designer's beliefs, intent and values; the logic of the user's context; and the logic of the system's context. A structuring chart for conclusions at the middle level of abstraction in Semiotic Analysis, similar to the one presented in Figure 1 can be used to organize the inspector's findings.

Keeping with the small example illustrated in Figure 2, the inspector will now explore facets of mediated designer-user communication that are not easily tractable by user-centered approaches. Since Semiotic Engineering contemplates the intentions and needs of *both* parties involved in mediated communication, interaction design *is* compatible with the **designer's** intent to advertise the benefits of purchasing QDA Miner's full version. This kind of communication is not necessarily compatible with the user's context, since users of the free version may not be able or willing to purchase the full version. Finally, regarding the logic of the system's context, the free version might consistently be viewed as an active piece of commercially-targeted communication. This stage of analysis reveals the power of 'persuasive technologies', extensively studied by BJ Fogg [36]. Notice that these conclusions rely on abstractions over *patterns of communication* (exposing the full version's static interface, but intercepting the user's access to functions that are only available if the user obtains a license to use it), which ultimately constitute an advertising strategy.

In the higher level of abstraction in the process of Semiotic Analysis, the inspector must evaluate overall structure and pragmatic adequacy of the computer-mediated social communication sent from the inspected artifact's producers to the consumers. This provides a solid basis to appreciate the (chances of) effective and efficient *metacommunication* carried out by the designer's proxy when users interact with the system. The factors listed in the structuring chart for conclusions at the higher level of abstraction in Semiotic Analysis, similar to the one presented in Figure 1, include not only the characterization of relations between **senders** and **receivers** of metacommunication (*e. g.* relations of power and control), but also the signs that are used to define the **context** of interaction. Additionally, at this stage of abstraction, inspectors can evaluate the **codes** and **channels** through which designer-user metacommunication is carried out.

To illustrate the conclusions of analysis at this final stage, we return to the small example in Figure 2. The inspector can now refine the analysis of how QDA Miner Lite achieves a company's *negotiation* with its customers by means of a limited functioning version of the product it wants to sell, a common practice in the software industry. The kinds of communicability features associated with the above-mentioned factors begin with the verification that system's creators explicitly participate in communication, using persuasive discourse (they take full advantage of their role of communication senders). The receivers of these senders' message may however not be the targets that senders want to reach (our profiled user is not likely to be in a position to invest a considerable amount of money to purchase this tool for introductory class exercises with qualitative data analysis). Therefore, whereas interaction may be the right context for commercial negotiations in some cases, in some others they may not. The code used in this instance of communication is perfectly understandable, and the channels of communication allow senders to reach the all users easily. However, to *talk back* to the senders declining this sort of commercial mediated commercial conversation, users should know how to turn of this *default* communicability feature of QDA Miner Lite. Curiously, this is done with one of the Help menu options (Figure 3).

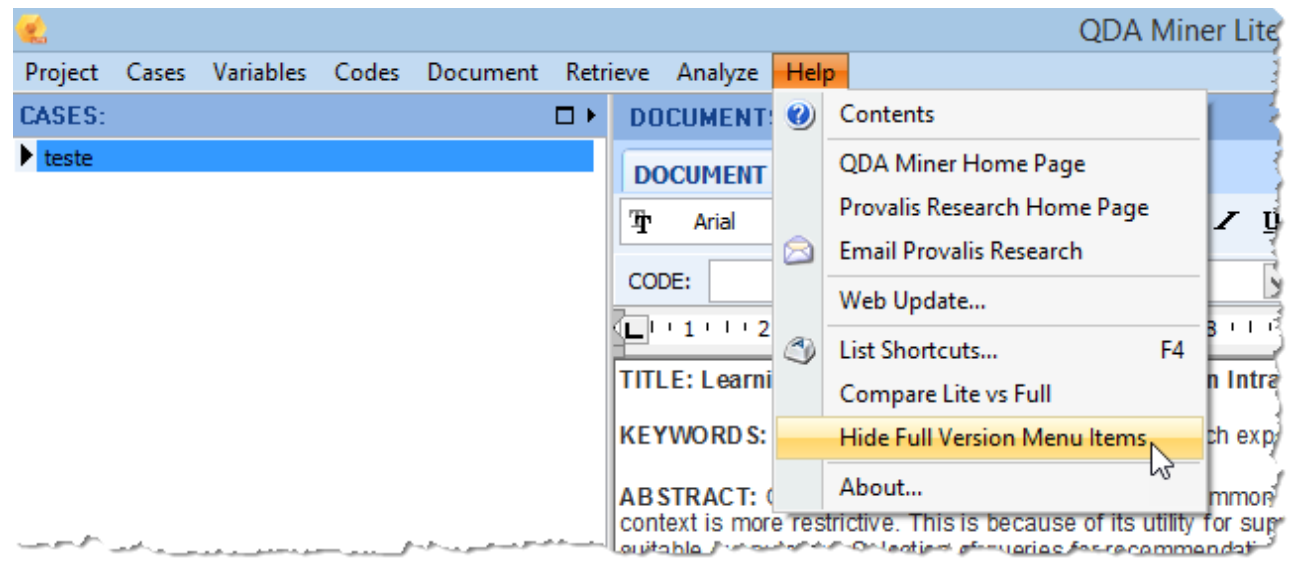

**Figure 3: Hiding Commercially-Oriented Communication**

The conclusion of semiotic analysis of this small example, as is typical of interpretive methods, is the result of iterated meaning associations and inferences drawn from them, as well as from iterated revisions of previous interpretations and reasoning. Inspectors will navigate from lower to higher levels of abstraction, and then revisit these levels and enrich the analysis with questions and findings that are triggered by the interim conclusions that such abstractions help him to achieve. To illustrate the effects of iterations, including commercial advertisement in the flow of interaction by default, and placing the control to turn it of as an option (among other commercially-oriented options) of the Help menu (Figure 3) is significant. Should users not realize that they can change the default configuration, they could be annoyed by the persisting advertising line of communication. It would be pragmatically more appropriate to let users decline this sort of conversation when they first meet it. In Figure 4, we show the sketch of a dialog to replace the one in Figure 2.

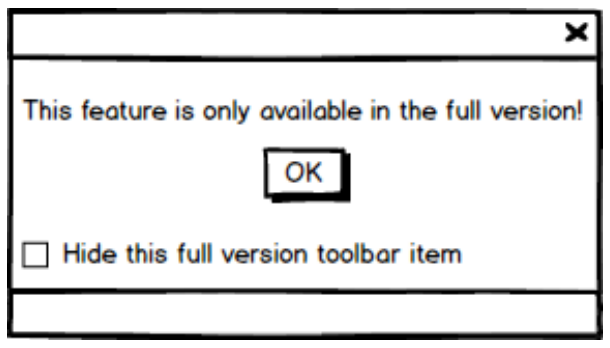

**Figure 4: A redesigned dialog to change default configuration**

**Cognitive Analysis**, in turn, is achieved in two steps. They correspond to the two final analytical steps defined by Blackwell and Green [1]. The preceding steps defined by the authors are accounted for in **SigniFYI-CDN** by the procedures defined for the **Preparation** block. Cognitive Analysis thus amounts to the following:

Given the specified user profile and inspection scenario, for every task that the user must or might reasonably wish to do, the inspector must:

1. Examine the cognitive dimensions of the notations that the user must understand and use (Figure 5);
2. Decide if the cognitive loads required for achieving the task are appropriate.

| Cognitive Dimension | Description |
|---|---|
| Abstraction | Types and availability of abstraction mechanisms |
| Closeness of Mapping | Closeness of representation to domain |
| Consistency | Similar semantics are expressed in similar syntactic forms |
| Diffuseness | Verbosity of language |
| Error-Proneness | The notation invites mistakes and the system gives little protection |
| Hard Mental Operations | High demand on cognitive resources |
| Hidden Dependencies | Relevant relations between entities are not visible |
| Premature Commitment | Constraints on the order of doing things |
| Progressive Evaluation | Work-to-date can be checked at any time |
| Provisionality | Degree of commitment to actions or marks |
| Role-Expressiveness | The purpose of an entity is readily inferred |
| Secondary Notation | Extra information in means other than formal syntax |
| Viscosity | Resistance to change |
| Visibility | Ability to view entities easily |

**Figure 5. The Cognitive Dimensions of Notations**

As suggested by their names (Figure 5), the presence or absence of cognitive dimensions can have positive or negative usability effects. To illustrate aspects of the CDN analysis with an example related to the one we used for the Semiotic Analysis, we take the toolbar in Figure 2 (magnified in Figure 6) where there is an icon representing a camera ('take a snapshot of the table'). Copy to clipboard functionality is fully available in QDA Miner Lite, which means that the user may *change the strategy to compose his report* – take a snapshot of QDA Miner Lite elements that he wishes to include in the report and then paste it in the report document created with some other tool. This feature shows that interaction to use code frequency tables in presentations of qualitative analysis is *viscous* in QDA Miner Lite. The artifact offers *resistance* to achieve the user's goal in alternative ways (see the definition of 'Viscosity' in Figure 5).

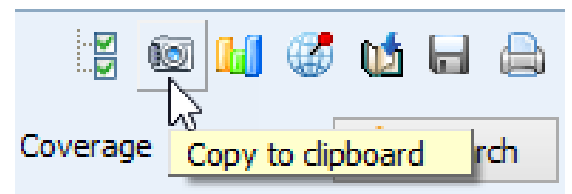

**Figure 6. QDA Miner Lite's snapshot tool**

Likewise, the fact that clicking on the report tool (Figure 2) triggers a commercial message rather than the announced functionality is a problem of 'Role Expressiveness'. The purpose of keeping that function active in QDA Miner Lite is not clear to the user. In fact, in terms of usability, this is additionally a case of 'Error-Proneness'. Users are likely to be misled by the icon and its tooltip, and click on it with the purpose of *adding the code frequency table to the report*.

**Articulation**, the last procedural step of the **SigniFYI-CDN** method allows the inspector to consider the interplay of social and psychological aspects of interaction. The semiotic and cognitive dimensions can be mapped to a *grid* structure (Figure 7), where he can plot the relations that he sees among them. By so doing, the inspector *articulates* the integrated semio-cognitive analysis of the inspected artifact.

Articulation is intensively iterative and typically produces multiple grids (*e. g.* one for each design feature or issue under analysis), although different inspectors may use the grid in different ways. The process is concluded when the inspectors have built a logical *argumentation* supporting their judgments about the communicability and usability of the inspected artifact's interaction design. The argumentation is grounded on interactive and metacommunication evidence collected throughout the analysis.

In Figure 7, we illustrate relations between communicability and usability issues associated with the example explored in this section. The inspection should recognize how communicability considerations modulate the significance of usability issues. For example, four cognitive dimensions – 'Closeness of Mapping', 'Error-Proneness', 'Role Expressiveness' and 'Viscosity' are associated with semiotic dimensions at different levels of abstraction. Whereas at the lower level, as already mentioned, the semiotic and cognitive analysis reciprocate results, focusing on how difficult it may be for the user to understand or decide what to do with the system, at higher levels of abstraction the power of intentionally provocative interactive discourse to persuade the user to acquire the product's license leads the inspector to conclude that designers may wish to cause usability problems for users, in an attempt to fulfill the company's commercial objective. This is a good example of what the precedence of Semiotic Analysis over Cognitive Analysis in **SigniFYI-CDN** procedural steps means.

Likewise, the inspector may ponder that facilitating the change of default values by communication like the one suggested in the redesigned sketch in Figure 4 would only make sense if QDA Miner Lite designers would agree to present the artifact in two ways: a working piece of demonstration and advertisement of QDA Miner Full Version's features; and a simpler, usable, useful and even enjoyable version of QDA Miner that can be used in teaching, learning and modest research project contexts.

## HIGHLIGHTS OF THE EVALUATION OF QDA MINER LITE

In this section we highlight additional aspects of **SigniFYI–CDN** analysis. The section does not present a complete evaluation summary, and should not be taken as a verdict on the quality of interaction with QDA Miner Lite. We use a new inspection baseline. The **profiled user** is Cynthia, a young Computer Science researcher, who has done a fair amount of qualitative data analysis with discourse and content analysis techniques. She has been using text editors and spreadsheets as support tools, but now she will be using QDA Miner Lite v. 2.0.4 [10] for the first time. She collaborates in a project with two other colleagues, who have watched a tutorial on YouTube and found that QDA Miner Lite can speed up their work, especially in subsequent phases of their joint project.

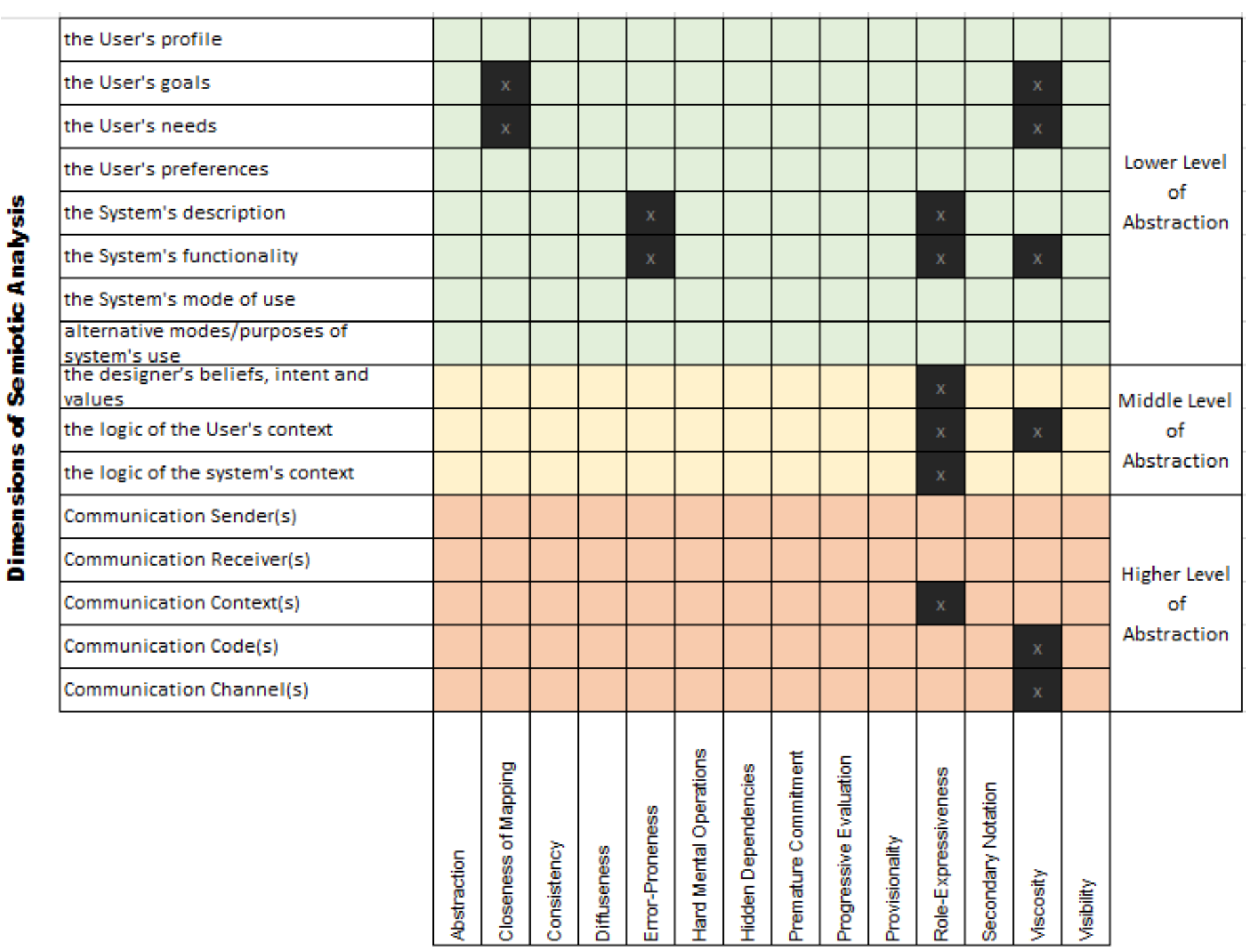

**Figure 7. An instance of the Articulation Grid**

The **inspection scenario** defines that the group is currently doing a systematic literature review (SLR) of Computer Science publications that are not classified as HCI publications. Their aim is to see how *humans* are viewed in this kind of literature. At this stage they have hundreds of citations and abstracts to examine. Their unit of analysis is a 'title+keywords+abstract' (TKA) triplet representing each publication. The data is stored in a single RTF file (Figure 8), listing all TKA triplets in the current collection. Each member of the group will do their *coding* (*i. e.* the classification and categorization) of text spans with relevant content for the analysis. They will then exchange QDA Miner Lite projects and discuss each other's analysis to validate their individual strategies of analysis and consolidate conclusions.

As the very name of the activity suggests, *coding* in qualitative data analysis is a classic notation-intensive task, hence the interest of using QDA Miner Lite in this illustration. For lack of space, we will concentrate our illustration on design features related to dataset formats.

Online help content explicitly communicates that:

"A case is the basic unit of analysis of a project. It typically represents an individual, an organization, or a group. A case can contain several documents as well as numerous […] variables."

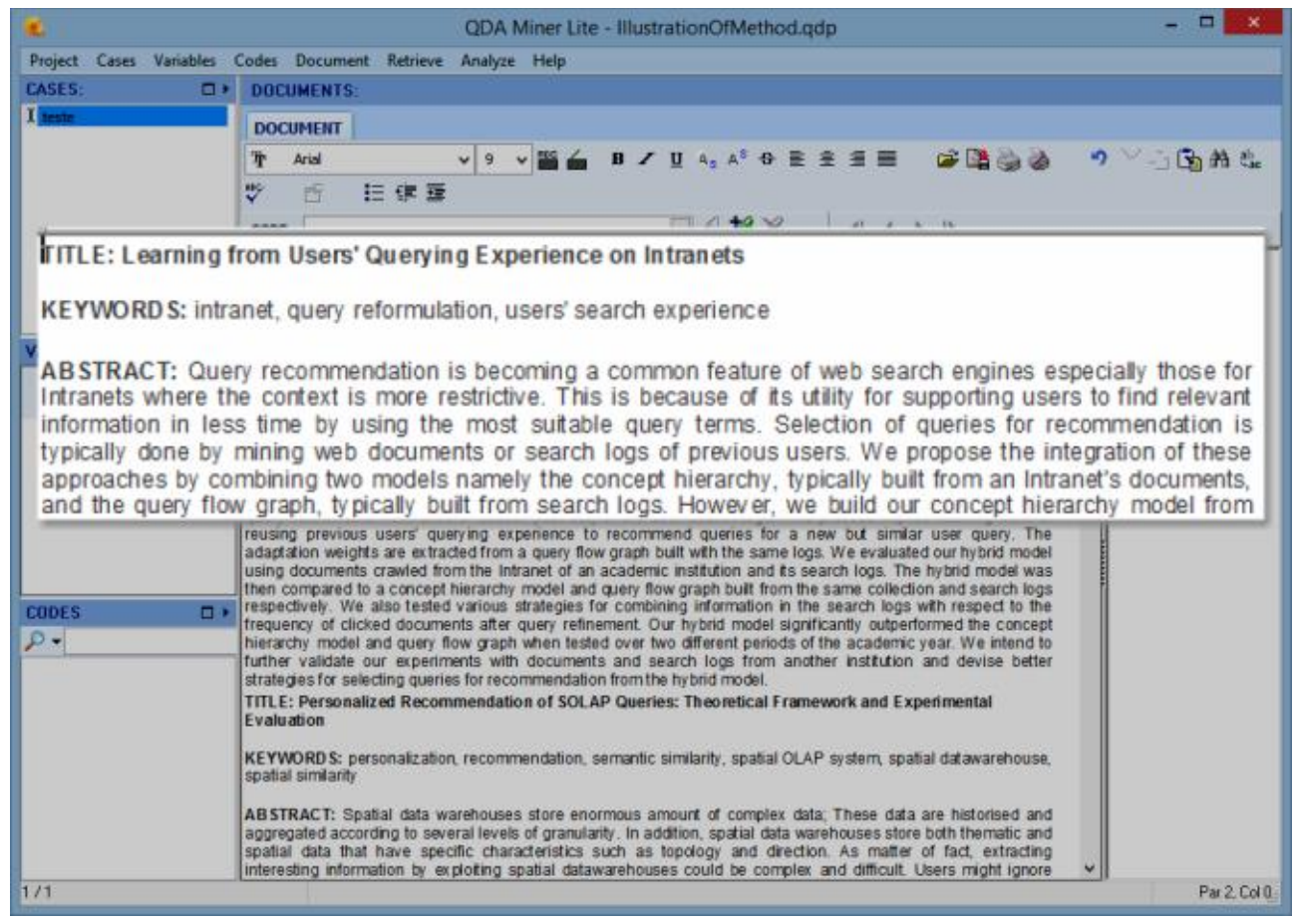

**Figure 8. QDA Miner Lite interface when importing a single RTF file**

According to the scenario, Cynthia's unit of analysis is the TKA triplet. However, the current format of her data is a single RTF file with all triplets, which constitutes the only *case* and *unit of analysis* that QDA Miner Lite 'sees' in her project (Figure 8). Cynthia therefore experiences an early communicability problem: the system (on behalf of its designers) *tells* her that the entire RTF file is the unit of analysis that she is working with. At this stage, she does not know how to tell the system that the RTF file is actually the *container* of hundreds of units of analysis. The inspector initially relates this problem to the following semiotic dimensions, at the lowest level of abstraction: user's goals (she wants to work with TKA triplets, not the entire text in the file); user's needs (she cannot create another unit of analysis for the current file); system functionality (the system does not support Cynthia's interpretation of the task she has to do). By interleaving semiotic and cognitive analyses, the inspector associates the communicability problem with the following cognitive dimensions: 'Closeness of Mapping' (the *system's*, actually the *designer's*, representation of Cynthia's task does not match her mental model); 'Premature Commitment' (the representation of data, even before a QDA Miner Lite project is created to analyze it, has a huge impact on the process of analysis); and 'Role Expressiveness' (the role of documents in project's *cases* is not readily inferred when *novices* begin to use QDA Miner Lite).

By navigating the articulation grid vertically (across different levels of abstraction in semiotic analysis) and horizontally (across different dimensions in cognitive analysis), the inspector plots his interpretations on the grid (see Figure 7 for an example of what articulation grids look like at this stage). Prompted by the relations he sees on the grid, the inspector infers that the high level message jointly conveyed by online help content, static interface signs and dynamic interaction possibilities is that QDA Miner Lite requires more preliminary knowledge than *novices* like Cynthia typically have. Only users who are familiar with how QDA Miner Lite and/or similar tools work would be able to escape the 'Premature Commitment' and 'Role Expressiveness' problems verified with the inspection. This partial articulation also communicates important aspects of the designers' beliefs regarding novice users. 'Premature Commitments' and 'Role Expressiveness' problems can be resolved when novice users learn (a potential 'Hard Mental Operations' problem) the logic of the system's design in one of two ways: trial and error; or instructed interaction. Trial and error amounts to doing what Cynthia has done so far: take *the wrong way* first, and backtrack when an 'error' is found. Instructed interaction amounts to start by reading guides, instructions and documentation before jumping into interaction. Yet, if the inspector examines (as he should) this alternative path, that is, if Cynthia carefully reads documentation before she begins to use QDA Miner Lite, he should realize that Cynthia would get a clear and direct communication from designers precisely about the kind of project that she has to do. Online documentation states that QDA Miner Lite can import RIS files (a commonly used format for bibliographic databases), and shows how to do it.

"QDA Miner allows you to directly import data files from spreadsheet and database applications, as well as from plain ASCII data files (comma or tab delimited text). The program can read data stored in the following file formats:

- ...
- *Reference Information System (RIS) data files"*

Following this alternative, Cynthia can import her dataset in RIS format, which in contrast with the importation of the RTF file shown in Figure 8, produces a very different effect shown in Figure 9. Now, Cynthia's project has hundreds of cases, each one shown in tabbed displays (see "Title", "Abstract", "Keywords" in the magnified "Documents" area that corresponds to the selected Case #1 in the "Cases" area).

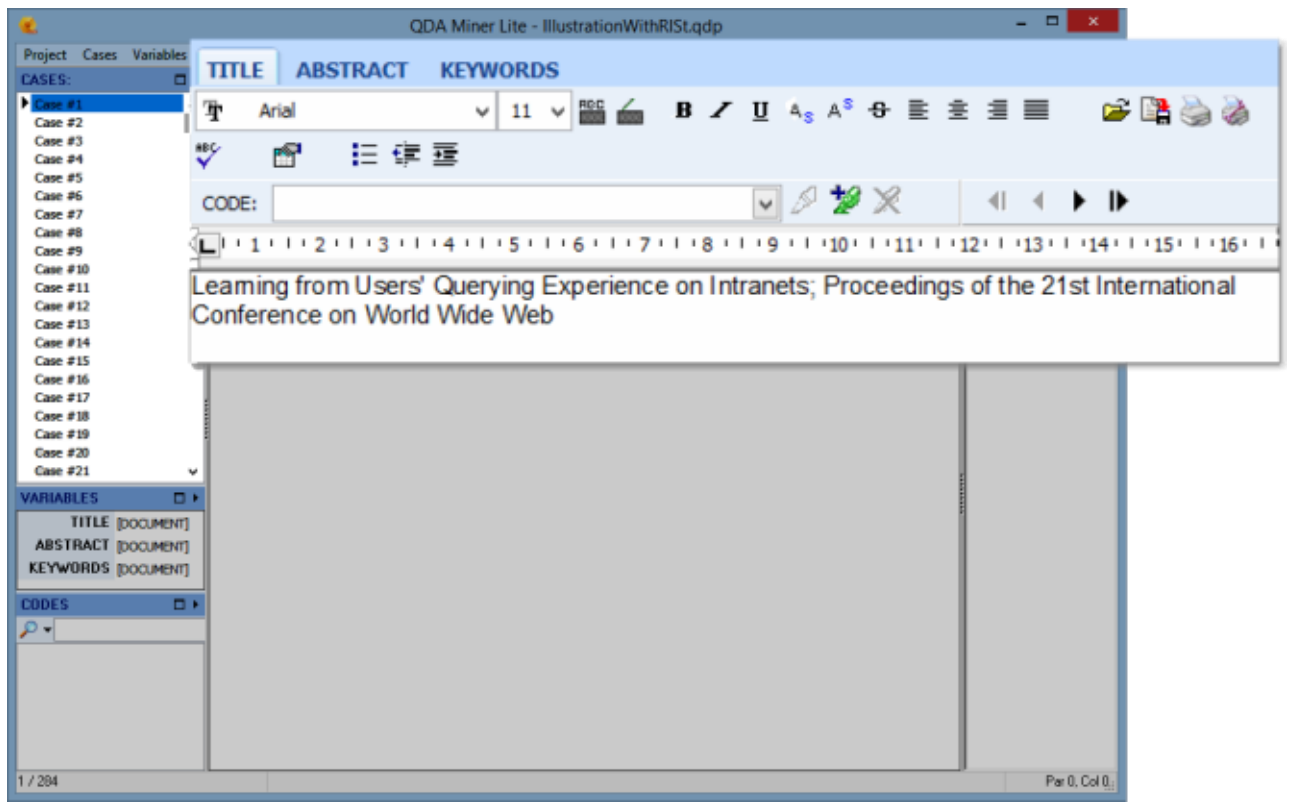

**Figure 9. QDA Miner Lite interface when importing a single RIS file**

The interleaved analysis allows the inspector to identify usability issues with this option, too. Cynthia cannot see (a 'Visibility' problem) the entire unit of analysis that she is using in her project. The *tabbed view* forces her to see the elements of the TKA triplet one at a time. Moreover, this visualization strategy adopted by QDA Miner Lite designers cause an additional cognitive problem, which CDN characterizes as 'Diffuseness'. The notation for the user's conceptual *unit* now extends over *multiple* representations. Additionally, working with this alternative, the user has to perform 'Hard Mental Operations' to keep her conceptual unit of analysis mentally integrated while the interface communicates about it in segments that are not jointly visible.

In terms of communicability, the designer's communication to users is that QDA Miner Lite explicitly supports the qualitative analysis of bibliographic datasets (RIS files can be easily imported into a project). However, the effects of the dataset's structure on the user's conceptual framing of her activity can be considerable.

To appreciate the impact of this communicability feature on the user, the inspector should explore different coding strategies that could be used with the two alternative dataset structures, the RTF file's and the RIS file's. We illustrate what happens with one such strategy.

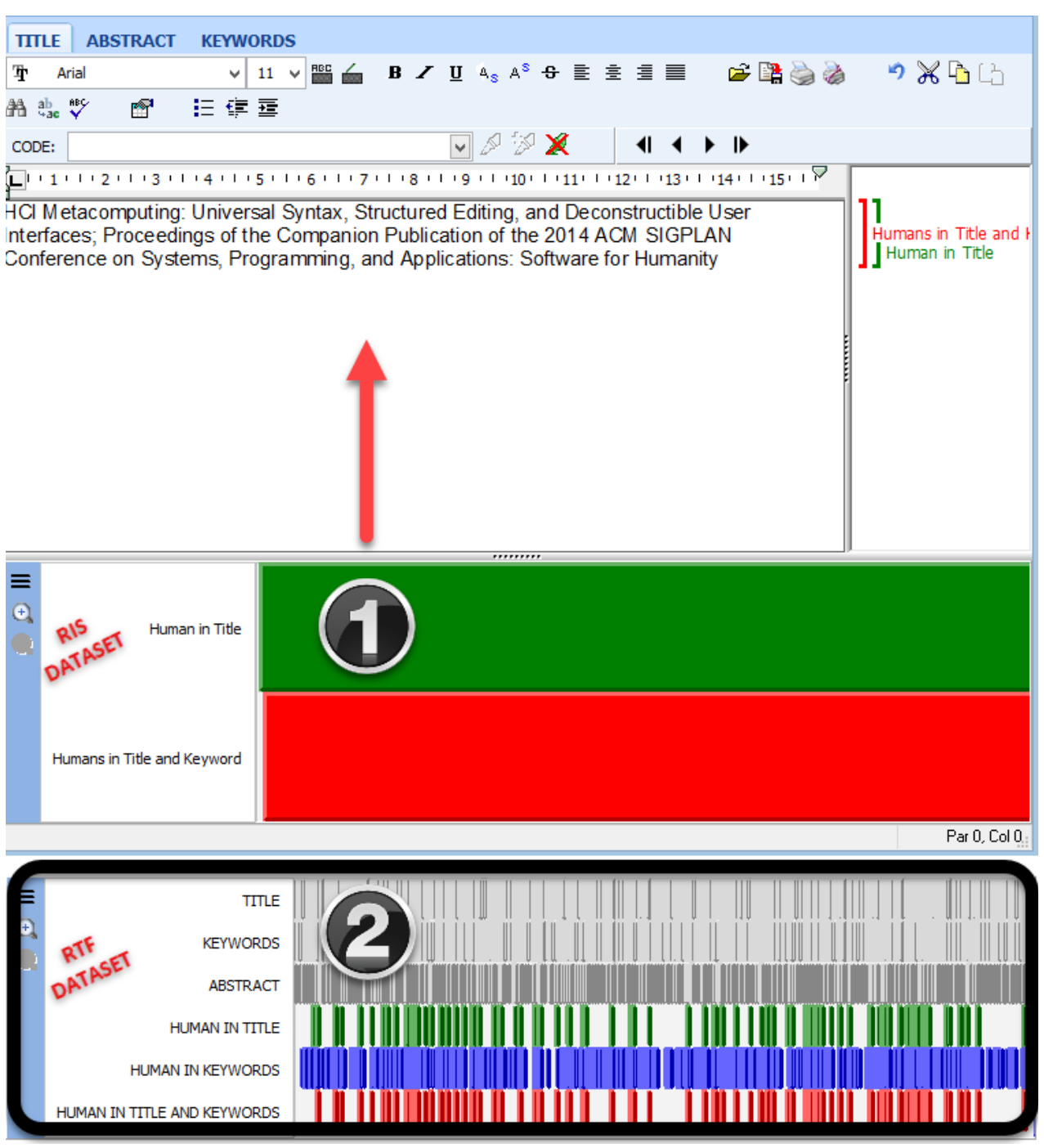

**Figure 10. An overview of coding with RIS (1) and RTF (2) Datasets**

Suppose that Cynthia begins her work by doing a 'rough coding' of data. She will subsequently analyze it in much more detail, to confirm or disconfirm her 'rough' intuitions. Her plan is: (1) to search items whose *title* include words referring to people (*e.g.* human(s), person(s), people, user(s), etc.) and *code them* with 'Humans in Title'; (2) to search items whose *keywords* include words referring to people and *code them* with 'Humans in Keywords'; and (3) to search items coded with both 'Humans in Title' and 'Humans in Keywords' and *code them* with 'Humans in Title and Keywords'. Her assumption is that items coded in step (3) correspond to a small fraction of the entire dataset and that these are strong candidates for her targeted set at this stage, publications that report on research focused on *people*.

The RIS file dataset allows her to carry out simple searches on the value of Title and Keyword fields, and to code them in batch. However, at first this strategy does not seem to work with the RTF file dataset. Although the file is visually *formatted* in TKA triplets (see Figure 8), it is not syntactically *structured* in the same way. In fact, to the system, it has no structure at all. Therefore, Cynthia's only chance to work with the RTF file is to structure the file herself, by means of 'Secondary Notation' operations. This line of interaction imposes considerable cognitive loads to Cynthia, but as a Computer Science graduate, she is used to working with notations.

Because abstracts and keywords are preceded by fixed strings of text, Cynthia can use automatic text retrieval functionality to search all paragraphs containing the strings "Title:", "Keywords:" and "Abstract:" and automatically code retrieved content with the corresponding Title, Keywords, Abstract *codes*. She thus *encodes* structure in the text file and can proceed with her strategy. Note that the strategy is possible not only at the expense of 'Secondary Notation' mental loads, but also of 'Hard Mental Operations' required for designing the structure, deciding how to represent it, and finally associating it to the appropriate file segments.

In terms of communicability, the inspector will find that, so far, QDA Miner Lite is *saying* that the recommended way to work with bibliographic datasets is to use RIS files. None of the complex structuring steps listed in the paragraph above are necessary. However, a very powerful communication casting serious doubts on this conclusion is reached when Cynthia tries to visualize the result of her coding in order to get an intuition of what her strategy may *mean* – a critically important step throughout the entire process of qualitative data analysis.

In Figure 10 we compare what QDA Miner Lite interface communicates to Cynthia in each case. With the RIS file Dataset (number '1'), the "Coding Overview" function shows all coding contained in a particular tab of a particular case (*e.g.* all codes in the Title of a particular publication). In Cynthia's context, this visualization is of no use, especially because the same information is already clearly communicated by the coding visualization to the right of the textual content (follow the red arrow in the image). The "overview" cannot see through structural borders that QDA Miner Lite recognizes in the dataset format. In comparison, the visualization produced for the RTF file Dataset (number '2', at the bottom of Figure 10) has superior communicability, usability and utility. With the superimposed structure, the RTF Dataset allows Cynthia to intuit quickly where her strategy is leading her. The first three rows corresponding to the structuring codes that she used show that the codes have been applied to all data items (as they should). The same is true for the 'Human in Keywords' row, which *tells her* that this particular code, in itself, is not informative now. The other two code rows – 'Humans in Title' and 'Humans in Title and Keywords' – have gaps in the dataset, although they visually suggest that they have been applied to the same data items, or nearly so. With this communication, Cynthia can quickly conclude that probably the only relevant content-related *rough* coding that she should keep is 'Human in Title'. Cynthia's initial strategy has been less productive than she expected, which should lead her to experiment other strategies.

The inspector can use the articulation grid once again to see the interplay of communicability and usability issues with the coding overview function, depending on which dataset format is being used. The design tradeoffs are clear. If Cynthia follows the designers' suggestions *by the book* and chooses the RIS Dataset, she will be spared the trouble of designing and implementing structure on what QDA Miner Lite sees as a flat file. However, the counterpart of this positive aspect is a questionable 'Closeness of Mapping' with the object of analysis that she has mentally construed. The separation of data items into individual *structured cases* introduces undesirable cognitive complexity (*e. g.* 'Hard Mental Operations', 'Diffuseness') that, in apparent contradiction with the message that SLRs figure among QDA Miner Lite's targeted types of qualitative data analysis, communicates that SRL analytic processes are not as smoothly supported as the automatic importation of RIS datasets suggested that they would be.

Interestingly, messages about functionality that is only available in QDA Miner Full Version do not suggest the illustrated communicability and usability problems are exclusive to the Lite, free version. An inspection of disabled or intercepted functionality calls, whose meanings are communicated only by their *names* (see Figure 2), hasn't identified any communication that SRLs are better supported by QDA Miner's Full Version than they are in the free version.

Taking the next step up the abstraction scale, the inspector should conclude that QDA Miner Lite metacommunication does not include efficient and effective messages about coding *strategies*. The designers' messages are centered on coding *operations*. Cynthia had to use her own mental resources to create a solution to be able to work with the dataset of RTF file. Moreover, the outcome of all of her creative effort is not readily available for ulterior *reuse*, by herself or her colleagues. QDA Miner Lite provides no scripting facilities. Therefore, if she faces the need to use a similar strategy in the future, she will have to go through the entire process again. If the designers had taken a different perspective on qualitative data analysis, paying as much attention to the *process* of coding (*e. g.* supporting the elaboration, codification, execution and reuse of several *coding strategies*) as they paid to the *product* of coding (which supports the creation and many additional manipulations and operations on *codes*), some of the arguments against the use of CAQDAS coming from qualitative data researchers might change. According to Gibbs [37], "there has been a long lasting debate about the role of coding in [qualitative data analysis (QDA)] and *a fortiori* in CAQDAS." While for some coding is "just a matter of data management", for others "coding [...] requires skilled perception and artful transformation" and is part of the core theory-building process in QDA. Gibbs's description of CAQDAS is, however, a powerful evidence of what the underutilization of software as a social means of communication can do. In his words, "For the programs, coding is simply a process of attaching a name or tag to a passage of text or an area of an image or a section of a video or audio recording. The software does not care about the analyst's motivation for this act of tagging and it certainly does not understand any interpretation given to it." [37] While the anthropomorphization of *software* confirms the perspective that users engage in human-like communication with systems, the perception that *software* doesn't care about how researchers view their work testifies to the potential of progress that combined communicability and usability analysis and design can bring to the field.

## CONCLUDING REMARKS

We presented and illustrated a new inspection method called **SigniFYI-CDN**. With its roots in Semiotic Engineering [3,7] and the Cognitive Dimensions of Notations framework [1], it is well equipped to evaluate notation-intensive interaction, which has become critical in the presence technologies using massive volumes of data to learn and do things that until recently seemed to be exclusively human capabilities. Notations used in sophisticated data analytics applications and machine learning play a central role in algorithmic processing. Thus, as they come within reach of *end users*, intelligent assistants and recommendation systems whose behavior is driven by huge volumes of data *produced* by these *end users* call for appropriate interaction evaluation methods.

Large volumes of data that drive new technologies often come from different sources. They are created by different people or, if by a single person, they originate from many different contexts, over time. Hence, multiple notations can be (and typically are) used, making room for multiple situated interpretations. Our examples with QDA Miner Lite show how **SigniFYI-CDN** can capture the variation of notations meant to communicate similar meanings, as well as the variation of meanings that can be assigned to notations in different contexts. We thus expect that **SigniFYI-CDN** will be attractive to researchers interested in various kinds of notation-intensive interaction.

Our future work will explore two lines of investigation to consolidate the method. The first is to test **SigniFYI-CDN**'s performance with several technologies, in summative evaluation settings, where evaluators can extensively examine an achieved product. **SigniFYI-CDN** should also be productive in formative evaluation settings, but we plan to test this possibility after later. Once our method is improved and consolidated, the second line of investigation we will explore is to propose the integration of **SigniFYI-CDN** to de Souza and co-authors' SigniFYI Suite, as an addition that can serve as an introduction to the more fine-grained and sophisticated methods of semio-cognitive analysis in the suite.

## ACKNOWLEDGMENTS

Clarisse S. de Souza thanks CNPq, the Brazilian National Council for Scientific and Technological Development, for partially supporting this research (Grant 304224/2017-0).